\documentclass[hidelinks,onefignum,onetabnum]{siamart220329}
\usepackage{amssymb}
\usepackage{amsmath}
\usepackage{color}
\usepackage{graphicx}
\usepackage{hyperref}

\usepackage{tikz}
\usetikzlibrary{arrows.meta}
\usetikzlibrary{decorations.markings}
\usetikzlibrary {math}

\begin{document}
	
	\begin{center}
		{\large\bf
			%Modeling hysteresis: first principles vs phenomenological approach
			Hysteresis resulting from Lennard-Jones interactions
		}\\~\\
		Dmitrii Rachinskii$^1$, Andrei Zagvozdkin$^1$, Oleg Gendelman$^2$
	\end{center}
	
	\begin{center}
		$^1$Department of Mathematical Sciences, University of Texas at Dallas, USA\\
		
		$^2$Faculty of Mechanical Engineering, Technion---Israel Institute of Technology
		
	\end{center}
	
	\begin{abstract}
		The fundamental mechanism of hysteresis in the quasistatic limit of multi-stable systems is associated with transitions of the system from one local minimum of the potential energy to another. In this scenario, as system parameters are (quasistatically) varied, the transition is prompted when a saddle-node bifurcation eliminates the minimum where the system resides in. The objective of the present work is to specify this generic mechanism for systems of interacting particles assuming a natural single-well (Lennard-Jones) interaction potential for each pair of particles. We show multi-stability and present details of hysteresis scenarios with the associated bifurcations and transitions in a case study of constrained four-degrees-of-freedom four particle systems on the plane.
	\end{abstract}
	
	
	
	\bigskip
	
	\section{Introduction}\label{introduction}
	
	In this work, we revisit fundamentals of hysteresis modeling.
	
	Phenomenological models of hysteresis, which describe experimentally observed complex constitutive relations of materials and media, are ubiquitous in engineering
	%As such, they are
	and quite diverse.
	Examples include models of a stress-strain constitutive relation in elastoplastic materials (e.g.\ Prandtl's elastic-ideally plastic element \cite{prandtl}; Prandtl-Ishlinskii hysteresis model and its generalizations \cite{ishlinskii}; Moreau's sweeping process \cite{moreau}; rate-independent yield criteria \cite{lamba, prager}; Armstrong-Frederick \cite{af}, Chabo\-che \cite{chaboche}, Mroz nonlinear hardening rules \cite{mroz});
	related models of dry friction and creep-fatigue damage counting (Max\-well-slip friction model \cite{maxwell}; rainflow-counting algorithm of calculating fatigue \cite{rainflow}); magnetizing field-magnetization constitutive laws of magnetic materials (Preisach independent domain model \cite{preisach}; Bouc-Wen, Jiles–Atherton, Sto\-ner–Wohlfarth models \cite{ja, sw, wen}; Krasno\-sel'skii-Pokrovskii and Mayer\-goyz-Friedman models \cite{kp, mayergoyz}); pressure-saturation constitutive equations for flows in porous media (Parlange and Mua\-lem hysteresis models \cite{okane, parlange}); coupling of mechanical, magneto-electric and temperature variables in smart materials such as piezoelectric and magnetostrictive materials, shape-memory alloys and shape-memory polymers, to mention a few.
	
	The aforementioned models are intrinsically meso- or macroscopic and usually are loosely related to the microstructure. From one side, it can be considered as a sort of advantage---media and systems with broad variety of the microstructures can exhibit similar hysteresis behavior and can be described by similar models. From the other side, one always encounters a problem of adequate attribution of specific parameters to the effective models. Arguably, the best way is to evaluate parameters from the ``first principles", i.e.\ starting from the potentials of interatomic interactions, or potential energy landscape. Unfortunately, such relationships are usually far beyond the reach. The dynamic hysteretic behavior is governed by an interplay of the structural modifications and dynamic dissipation. Each of these two intrinsic components in microscopic models is not understood at the level of quantitative predictions, except possibly for very few very simple models. The goal of this work is to explore 
	%what are 
	the  
	minimal requirements to the interaction potential %of the explored system to exhibit 
	that warrant 
	the hysteretic behavior. The dissipation is 
	%thus adopted 
	assumed to be overwhelming.
	%, and the system in fact 
	As such, the system is considered in a quasistatic response regime. This simplification leaves aside %many crucial 
	a number of important 
	dynamic features of the process \cite{optics, dyn1, dyn2, dyn3, dyn4, dyn5, dyn6}, but still leaves a hope to achieve a useful %initial 
	approximation to realistic dynamics.
	
\begin{figure}[h]
			\includegraphics*[width=1\columnwidth]{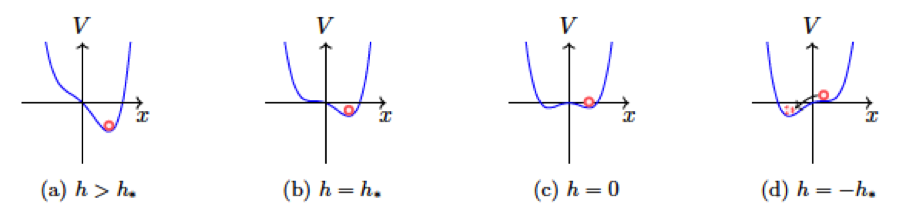}
		\caption{Particle in the double-well potential \eqref{V}.}\label{fb1}
\end{figure}

\begin{figure}[h]
		\centering
		\includegraphics*[width=.5\columnwidth]{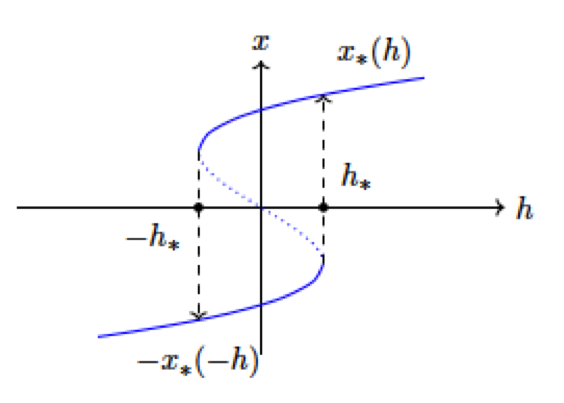}	
		\caption{Transitions of the particle from one to another (local) minimum of the potential energy shown in Figure \ref{fb1} as the exogenous field parameter $h$ (input) changes quasistatically. At the points $h=\pm h_*$, one minimum collides with the local maximum and disappears in a saddle-node bifurcation causing a transition to the other minimum. \label{fb2}}
	\end{figure}

	The fundamental mechanism of hysteresis associated with multi-stability and bifurcation is revealed by the following classical example (see e.g. \cite{visintin}). Let us consider a one-degree-of-freedom particle in the potential well
	\begin{equation}\label{V}
		V(x;h)= \frac{x^4}4 -\frac{x^2}2 - h x,
	\end{equation}
	where $h$ is the external field, which is varied quasistatically (i.e. the inertia is ignored).
	Assume that initially $h$ is large, so that $V$ has a unique minimum point $x=x_*(h)$ located on the positive semi-axis $x>0$, and the particle sits at this minimum, see Figure \ref{fb1}. Suppose that $h$ decreases.
	At the critical value $h_*=2/3\sqrt{3}$, the potential acquires the double well shape by developing the second minimum on the negative semi-axis $x<0$ through the saddle-node bifurcation, see Figure \ref{fb2}. At the critical value $h=-h_*$, the positive minimum is eliminated through the other saddle-node bifurcation, and the particle transitions to the remaining negative minimum $x=-x_*(-h)$. Next, assuming that from this point $h$ increases, the particle will be located at the negative minimum of $V$ until this minimum is eliminated through the saddle-node bifurcation at $h=h_*$, at which point the particle will transition back to the positive minimum $x=x_*(h)$, closing the hysteresis loop.
	
	This simple system displays important features of hysteresis.
	First, within the bi-stability range, $-h_* < h < h_*$, the state of the system (the position of the particle at the positive or negative minimum of the potential) is determined both by the concurrent and past values of the input $h$, hence one talks about history-dependence. Second, the history-dependence with the associated hysteresis loop manifests itself in the quasistatic limit of slow variations of $h$ (this fact is   referred to as rate-independence of hysteresis \cite{bs}). Third, each transition of the particle from one minimum of $V$ to another is associated with an irrecoverable energy loss.
	
\begin{figure}[h]
		\centering
\includegraphics*[width=1\columnwidth]{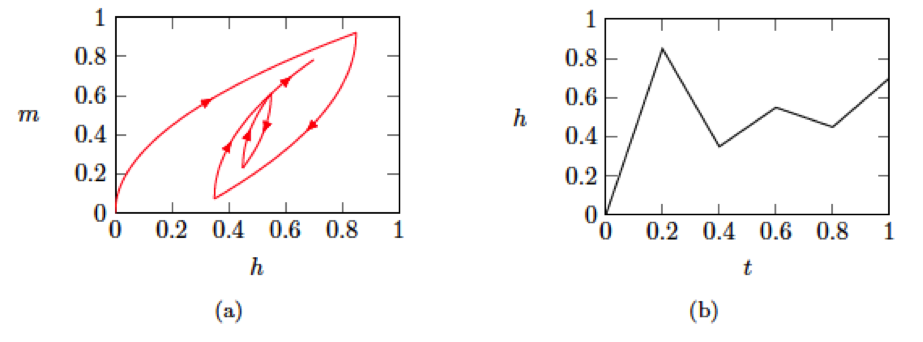}
		\caption{(a) A sample input $h=h(t)$ of the Preisach model. (b) Input-output diagram of the Preisach model depicting input $h$ (magnetizing field) vs output $m$ (magnetization) for the input shown on panel (a). The output is given by
			$
			m=\sum_{i=1}^N c_i\, \text{\rm sign}(x_i),
			$
			where $c_i$ are parameters (cf.\ \eqref{V}). The state $(x_1,\ldots,x_N)$ and output value $m$ at a given time $t$ depend both on the concurrent value of $h$ and a sequence of past extremum values of $h$, which are known as running main extremum values. }\label{fhm}
	\end{figure}

	%It is understood that the energy potential of a
	Generalizing the above example to
	multi-particle systems, the energy potential of a system with many degrees of freedom can have a large number of minimum points (metastable states). Further, as input variations cause the energy landscape to change, the same bifurcation mechanism (demonstrated by the double well potential) leads to a complex pattern of transitions between the states, creating a structure of hysteresis loops of the material constitutive law at a macrolevel. As one example, the Preisach model of magnetic hysteresis considers $N$ non-interacting particles, each in a double well potential \eqref{V}, i.e.\ the energy potential of the system is
	\[
	V(x_1,\ldots,x_N;h)=\sum_{i=1}^N \left( \frac{x_i^4}4-\frac{ x_i^2}2 - (h a_i + b_i) x_i \right),
	\]
	where $h$ is the input; $a_i, b_i$ are parameters. This potential has up to $2^N$ minima for a particular value of $h$, and %the model
	produces a specific structure of nested hysteresis loops (known as return-point memory), which are characterized by the so-called wiping-out and congruency properties \cite{mayergoyz}, see Figure \ref{fhm}. A hysteresis loop is an evidence that the system goes through one sequence of states as $h$ increases and then through a different sequence of states as $h$ decreases; or, that the system goes through the same sequence of states (in the reversed order) as $h$ decreases, but the transitions from one state to another and the reversed transitions occur at different values of $h$ (as in Figure \ref{fb2} in the case of the double well potential).
	
	As we see, in the Preisach model (and other phenomenological models of hysteresis phenomena), hysteresis of an individual particle is postulated. In this paper, we ask the following question:
	%whether
	{\it can hysteresis
		%can
		emerge  in a system of particles interacting via naturally non-hysteretic potentials?}
	To be more specific, we limit our discussion to systems of identical particles and the classical Lennard-Jones interaction.
	
	As a starting point, we make an observation that a chain of particles with the nearest neighbor interaction does not display hysteresis if the particles are elongated along a straight line (see the next section). Therefore, we look at systems of particles on a plane. As the main result, we answer affirmatively to the above question by presenting examples of simple 4-particle (constrained) planar configurations, which exhibit hysteresis. We provide a detailed analysis of the associated bifurcation scenarios (Section \ref{sec3}). The paper is concluded with a discussion of these results.
	
	\section{Preliminaries}\label{preliminaries}
	
\begin{figure}[h]
		\centering
\includegraphics*[width=.75\columnwidth]{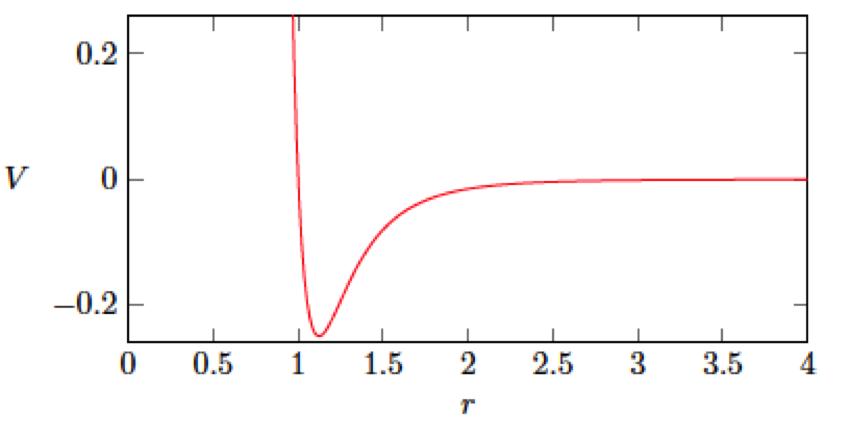}
		\caption{The  Lennard\,--\,Jones interaction potential of a pair of particles for \(\sigma = 1\). }\label{figure1}
	\end{figure}

	We consider a collection of $N$ particles in the potential field with the potential
	\begin{equation}\label{inter}
		V(\mathbf r;h)=V(\mathbf{r}_1,...,\mathbf r_N; h)= \sum_{1\le i <j\le N} \Phi_{ij}(r_{ij})+ h  \sum_{i}\mathbf a_i \cdot \mathbf r_i, \qquad \mathbf r =(\mathbf r_1,\dots,\mathbf r_N),
	\end{equation}
	where $\mathbf r_i$ is the position of the $i$-th particle; $h$ is a scalar input variable (such as the amplitude %of related to
	of external forcing, load, external field etc.); $r_{ij}=|\mathbf r_{i}-\mathbf r_j|$ is the Eucledian distance between the $i$-th and $j$-th particles; $\Phi_{ij}$ is the interaction potential of the pair of particles; $\mathbf a_i$ are vector-valued parameters; and, dot stays for the dot product.
	It is assumed that the two-particle interaction potential is the  Lennard-Jones potential
	\begin{equation}\label{vdw}
		\Phi_{ij}(r) = 4\varepsilon_{ij} \Phi_1(r), \qquad \Phi_1(r)= \left(\frac{\sigma}{r}\right)^{12} - \left( \frac{\sigma}{r}\right)^{6},
	\end{equation}
	see Figure \ref{figure1}.
	
	We consider the quasistatic evolution of the system in response to quasistatic variation of the parameter $h$. This evolution has intervals of (relatively) slow and fast dynamics. During the slow evolution, the system sits in a local minimum of the potential $V$, say $\mathbf r^-_*=\mathbf r^-_*(h)$, until this minimum point is eliminated via a saddle-node bifurcation as $h$ is varied. 
	The saddle-node bifurcation is the only generic mechanism creating/eliminating minimum points of $V$.
	
	At the bifurcation point $h=h_b$, the system %(instantaneously)
	transits to another minimum following the %(infinitely)
	fast antigradient dynamics
	\[
	\dot {\mathbf r} = -\nabla_{\mathbf r} V(\mathbf r; h_b), \qquad \mathbf r =(\mathbf r_1,\dots,\mathbf r_N),
	\]
	i.e.\ the system follows the one-dimensional unstable manifold of the saddle-node equilibrium point $\mathbf r_*^-(h_b)$ of the gradient field
	to another local minimum point $\mathbf r_*^+(h_b)$ of the potential. This transition is (infinitely) fast compared to the slow dynamics following a minimum of $V$. The antigradient transition dynamics ensures that
	\[
	\dot V = - |\nabla_{\mathbf r} V(\mathbf r; h_b)|^2,
	\]
	hence $V$ decreases along the transition trajectory, and
	\[
	V(\mathbf r^-_*; h_b)-V(\mathbf r^+_*; h_b)>0.
	\]
	This positive quantity represents an irreversible energy loss (dissipation) associated with the transition. If, after (several) transitions, $h$ returns to its initial value and the system returns to the same minimum of $V$ where it started from, then hysteresis is observed.
	
\begin{figure}[h]
		\centering
\includegraphics*[width=.8\columnwidth]{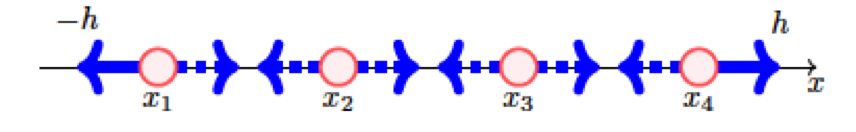}
		\caption{System of $N=4$ particles with nearest neighbor  interactions on a line. Each force (shown by an arrow) has amplitude $|h|$. \label{chain}}
	\end{figure}

	We start by showing that a one-dimensional chain of $N$ particles with nearest neighbor interactions and an external forcing applied at the ends of the chain does not exhibit hysteresis. Namely, let us consider the potential
	\[
	V(x_1,...,x_N; h)= h(x_1-x_N)+\sum_{i=1}^{N-1} \Phi_1({x_{i+1}-x_i}),
	\]
	where $x_1<\cdots <x_N$ are positions of the particles on a straight line,
	$\Phi_1$ is the Lennard-Jones two-particle interaction potential (cf.\ \eqref{vdw}), and the opposite forces $-h$ and $h$ are applied to the two particles at the ends of the chain, see Figure \ref{chain}. Using the variables $q_i=x_{i+1}-x_i$, the potential reads
	\[
	V(q_1,\ldots,q_{N-1};h)=\sum_{i=1}^{N-1} \bigl(\Phi_1(q_i)-h q_i\bigr).
	\]
	This potential does not have critical points for $h>h_*=126/169$ (in this case, the external force expanding the chain exceeds the maximal attraction force between the particles, and the chain breaks). For $0<h<h_*$, the potential has one local minimum and one local maximum point.
	For $h<0$, the minimum is global and is a unique critical point of $V$
	(the maximum disappears at infinity as $h$ becomes negative:
	$h>0$ corresponds to the expansion and $h<0$ to the contraction of the chain by the external forces). Since $V$ has at most one minimum, the system does not exhibit hysteresis.
	
	\section{Case study of two-dimensional structures}\label{sec3}
	
\begin{figure}[h]
\includegraphics*[width=1\columnwidth]{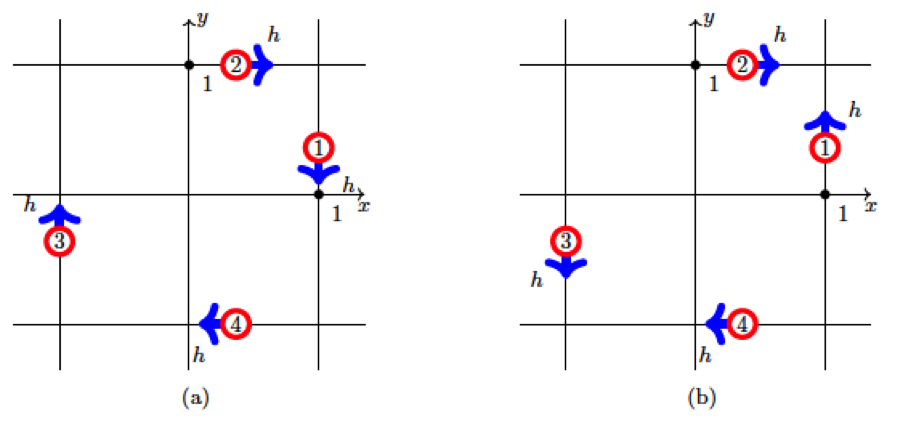}
		\caption{Four constrained particles with pairwise Lennard\,--\,Jones interaction under (a) rotational external forcing; (b) expansion. }\label{figure4:a}\label{figure4:b}\label{figure4}
	\end{figure}

	As a prototypical example of hysteresis
	in a system of particles interacting via the  Lennard-Jones potential,
	we consider the system of four identical particles shown in Figure \ref{figure4}.
	The particles are placed on the $(x,y)$-plane; the coordinates of the $i$-th particle are denoted by $(x_i,y_i)$.
	Particles 1 and 3 are constrained to the vertical lines $x=1$ and $x=-1$, respectively, while particles 2 and 4 are constrained to the horizontal lines $y=1$ and $y=-1$, i.e.\ each particle has one degree of freedom. We use the notation $q_1=y_1$, $q_2=x_2$, $q_3=y_3$, $q_4=x_4$ for the system coordinates. Assuming the Lennard\,--\,Jones pairwise interaction between the particles, the system potential is
	\begin{equation}\label{v0}
		V^{\sigma}_0(q_1,q_2,q_3,q_4)=\sum_{1\le i<j\le 4} \Phi_1(r_{ij})=\sum_{1\le i<j\le 4} \left(\left(\frac{\sigma}{r_{ij}}\right)^{12} - \left( \frac{\sigma}{r_{ij}}\right)^{6}\right),
	\end{equation}
	where $r_{ij}$ is the Eucledian distance between particles $i$ and $j$.
	
	We show hysteresis in this system under external forcing. Two types of forcing will be considered.
	
	As the first example, we will assume that particles 1\,--\,4 are acted upon by the constant forces $h$, $-h$, $-h$, $h$, respectively, along the lines of their motion as shown in Figure \ref{figure4:a}. In this case, the full potential of the forced system is 
	\begin{equation}\label{rot}
		V^{\sigma}_h(q_1,q_2,q_3,q_4)= h (q_1-q_2-q_3+q_4) + V^{\sigma}_0(q_1,q_2,q_3,q_4).
	\end{equation}
	This system is discussed in Section \ref{rotation}.
	
	Another example, in which the constant external forces acting on  particles 1 -- 4 are $h$, $h$, $-h$, $-h$, respectively, and
	and the corresponding potential is
	\begin{equation}\label{exp}
		\hat V^{\sigma}_h(q_1,q_2,q_3,q_4)= -h (q_1+q_2-q_3-q_4) + V^{\sigma}_0(q_1,q_2,q_3,q_4),
	\end{equation}
	is considered in Section \ref{expansion}, see Figure \ref{figure4:b}.
	
	\subsection{Unforced system}\label{unforced}
	Let us first discuss the unforced system with potential \eqref{v0}.
	We show that, for certain ranges of the parameter $\sigma$, this potential has multiple minimum points. In other words, for such $\sigma$, the system with potential  \eqref{rot} (resp.\ \eqref{exp}) is multi-stable when $h=0$, i.e.\ the external forcing is zero.
	
	Potential \eqref{v0} is invariant with respect to the action of the dihedral group $\mathbb{D}_4$ of symmetries of the square.
	A generating set of this group, consisting of the clockwise rotation $\rho$ by $\pi/2$ around the origin and the reflection $\kappa$ over the line $x=y$, acts on the configuration space of the system by mapping a point $\mathbf{q}=(q_1, q_2, q_3, q_4)$ to the points
	\[
	\rho (q_1, q_2, q_3, q_4) = (q_4,- q_1, q_2, -q_3), \qquad \kappa (q_1, q_2, q_3, q_4) = (q_2, q_1, q_4, q_3),
	\]
	respectively. We will use the subgroups $\mathbb{Z}_4=\{e, \rho, \rho^2, \rho^3\}$, $\mathbb{Z}_2=\{e, \kappa\}$  of $\mathbb{D}_4$.
	
\begin{figure}[h]
		\centering
	\includegraphics*[width=.7\columnwidth]{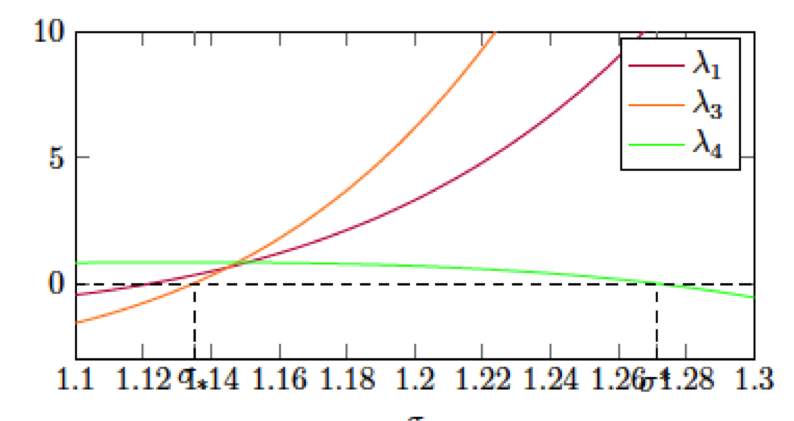}
		\caption{Eigenvalues $\lambda_1=\lambda_2$ {(magenta)} , $\lambda_3$ {(orange)}, $ \lambda_4$ {(green)} of the Hessian of potential \eqref{v0} at zero as functions of the parameter $\sigma$. %(a) 
			The eigenvalues are positive for $\sigma_*<\sigma<\sigma^*$.
		}\label{figure5}
	\end{figure}

	Let us consider the fully symmetric zero critical point $\mathbf{q}=0$ of the potential $V_0^\sigma$. A direct calculation shows that the eigenvalues of the Hessian of the potential at zero are
	\begin{equation}\label{lalala}
		\lambda_1=\lambda_2=\frac{9\sigma^6 (\sigma^6-2)}8, \quad \lambda_3=\frac{3 \sigma^6 (1663 \sigma^6-3552)}{2048}, \quad  \lambda_4=\frac{3\sigma^6 (544 - 129 \sigma^6)}{2048},
	\end{equation}
	see Figure \ref{figure5}. Hence, $\mathbf{q}=0$  is a (local) minimum point of the potential for the values of the parameter $\sigma$ from the interval
	\begin{equation}\label{stabilityint}
		(\sigma_*, \sigma^*)=\left(\left(\frac{3552}{1663}\right)^{1/6}, \left(\frac{544}{129}\right)^{1/6}\right)=(1.13483, 1.27107).
	\end{equation}
	At each end of the stability interval \eqref{stabilityint}, the anti-gradient field $-\nabla V_0^\sigma$ undergoes a supercritical symmetry braking bifurcation at $\mathbf{q}=0$.
	
	\medskip\noindent
	{\bf Symmetry breaking pitchfork bifurcation at $\sigma=\sigma^*$.}
	As $\sigma$ increases across the critical value $\sigma^*=1.27107$ where $\lambda_4 (\sigma^*)=0$, the anti-gradient field undergoes a supercritical pitchfork bifurcation producing a pair of minimum points
	\begin{equation} \label{qq}
		\mathbf{q}^*=(q^*, - q^*, - q^*, q^*), \qquad \kappa \mathbf{q}^*=-\mathbf{q}^*=(-q^*, q^*, q^*, -q^*)
	\end{equation}
	of the potential,
	which bifurcate from the critical point $\mathbf{q}=0$  as it changes stability and becomes a saddle.
	The pair of critical points \eqref{qq} exists for $\sigma>\sigma^*$, they form a $\mathbb{Z}_2$ orbit, and each of them is $\mathbb{Z}_4$-symmetric because $\rho \mathbf{q}^*=\mathbf{q}^*$.
	In other words, the pitchfork bifurcation at $\sigma=\sigma^*$ breaks the $\mathbb{Z}_2$-symmetry of the zero critical point but preserves the $\mathbb{Z}_4$-symmetry.
	
	\begin{figure}[h]
			\centering
		\includegraphics*[width=.6\columnwidth]{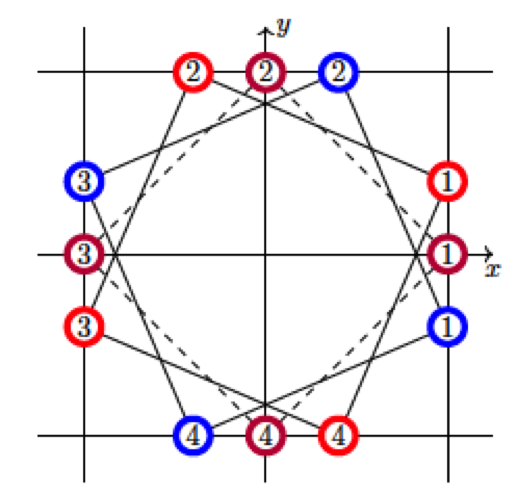}	
		\caption{Square shaped formations of the particles on the $(x,y)$-plane corresponding to minima \eqref{qq} of potential \eqref{v0} (red and blue) and the the square formation corresponding to the zero critical point (magenta).
			\label{figure65}}
	\end{figure}
	
\begin{figure}[h]
		\centering
			\includegraphics*[width=.8\columnwidth]{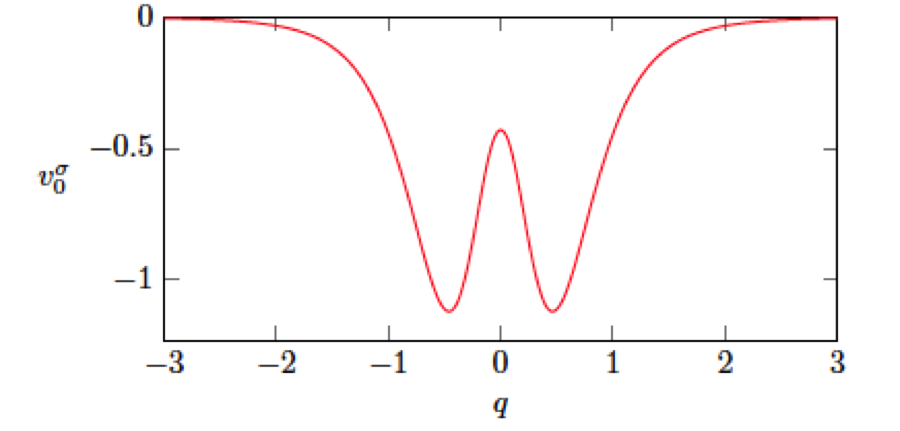}	
		\caption{The restriction $v^\sigma_0(q)$ of the potential $V_0^\sigma=V_0^\sigma(\mathbf q)$ to the one-dimensional subspace $L=\{ \mathbf{q}=(q,-q,-q,q), q\in \mathbb{R}\}$ of the configuration space
			for $\sigma=1.4>\sigma^*$. The subspace $L$ is invariant for the anti-gradient field and the action of the symmetry group $\mathbb{Z}_4$. \label{figure6}}
	\end{figure}

	It is important to observe that the one-dimensional subspace
	\begin{equation}\label{L}
		L=\{ \mathbf{q}=(q,-q,-q,q), q\in \mathbb{R}\}
	\end{equation}
	of points fixed by the symmetry group $\mathbb{Z}_4$ in the configuration space of the system is invariant for the anti-gradient flow, i.e.\
	$-\nabla V_0^\sigma(\mathbf{q})\in L$ for $\mathbf{q}\in L$. 
	Each point of $L$ in the configuration space corresponds to positioning of the particles in the corners
	of a square. In particular, the two squares corresponding to the minimum points $\pm \mathbf{q}^*$ of the potential are symmetric to each other
	with respect to the bisector line $x=y$, see Figure \ref{figure65}.
	Since $L$ contains the critical points $\pm \mathbf{q}^*$ given by \eqref{qq},
	these points can be found as minimum points of the restriction of $V_0^\sigma$ to $L$, which is given by
	\[
	v^\sigma_0(q)=V_0^\sigma(q,-q,-q,q)=\frac{ 129 \sigma^{12} - 1088 \sigma^6(1 + q^2)^3}{2048 (1 + q^2)^6},
	\]
	see Figure \ref{figure6}. In this way, one obtains
	\[
	q^*=\sqrt{\left(\frac{\sigma}{\sigma^*}\right)^2-1}
	\]
	for the minimum points $\pm q^*$ of the function $v^\sigma_0$ and for the components of minimum points \eqref{qq} of the potential $V^\sigma_0$.
	Further,
	by direct calculation, the eigenvectors of the Hessian at any point of $L$ are
	\begin{equation}\label{eigenvectors}
		(1, 1, 1, 1), \quad (1, -1, 1, -1), \quad (-1, -1, 1, 1), \quad (1, -1, -1, 1).
		% (1,0,1,0), \quad (0,1,01)
	\end{equation}
	Moreover, the corresponding eigenvalues at the critical points $\pm \mathbf{q}^*\in L$ of the potential equal
	\[
	\mu_1=\mu_2=\frac{1287 (\sigma^*)^{14}}{2176\, \sigma^2},
	\quad \mu_3= \frac{3 (\sigma^*)^{14}  (7373 (\sigma^*)^2- 397 \sigma^2)}{17408\,\sigma^{4} },
	\]
	\[
	\quad \mu_4=\frac{1161 (\sigma^*)^{14} (\sigma^2-(\sigma^*)^2)}{1024 \, \sigma^{4}}.
	\]
	Hence \eqref{qq} are minimum points of the potential for 
	\begin{equation} \label{addd}
		\sigma^*<\sigma<\sigma^{**}=\sqrt{7373/397} \sigma^*=5.47766.	
	\end{equation}
	
	At the point $\sigma=\sigma^{**}$, the minima $\pm \mathbf{q}^*$ destabilize in the direction $(-1,-1,1,1)$, which is perpendicular to $L$, and become saddles.
	
	In Section \ref{rotation}, we will consider the system with rotational forcing (potential \eqref{rot}) for $\sigma>\sigma^{*}$ and show hysteresis between the square shaped configurations of particles as the external forcing parameter $h$ is varied.	
	
\begin{figure}[h]
		\centering
	\includegraphics*[width=1\columnwidth]{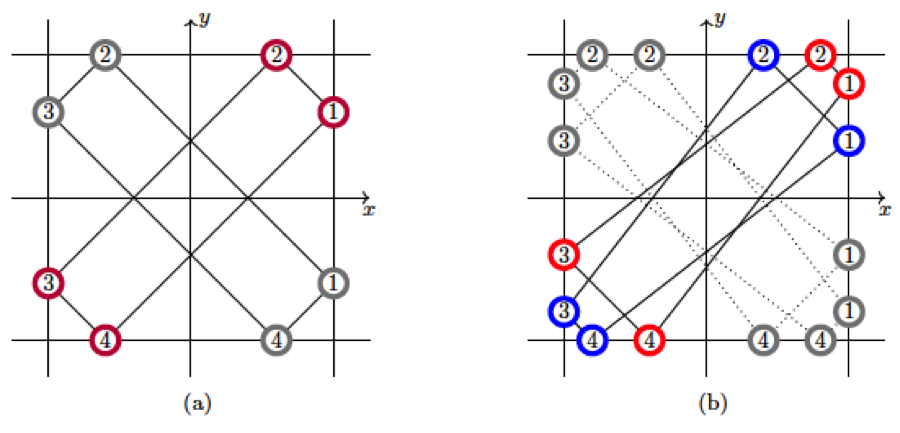}	
		\caption{(a) Rectangular formations of the particles on the $(x,y)$-plane corresponding to minima \eqref{qq1} of the potential $V_0^\sigma$.
			(b) Trapezoid shaped configurations of particles corresponding to minima \eqref{z4}.
			\label{figure66:a}
			\label{figure66:b}
			\label{figure66}}
	\end{figure}

	\medskip\noindent
	{\bf Symmetry breaking pitchfork bifurcation at $\sigma=\sigma_*$.}
	Now, let us consider the other bifurcation point, $\sigma=\sigma_*=1.13483$, where the zero $\mathbf{q}=0$ of the anti-gradient field loses stability.
	At this supercritical pitchfork bifurcation point, the additional (non-zero) critical points of $V_0^\sigma$ appear in a different anti-gradient flow invariant one-dimensional subspace,
	namely
	\[
	M=\{\mathbf{q}=(q,q,-q,-q), q\in \mathbb{R}\}
	\]
	(cf.\ \eqref{L}). More precisely, when the eigenvalue $\lambda_3(\sigma)$ (cf.\ \eqref{lalala}) crosses zero at $\sigma=\sigma_*$ as $\sigma$ decreases, see Figure \ref{figure5} (the orange line), the point $\mathbf{q}=0$ becomes a saddle, and a pair of minimum points of $V_0^\sigma$ forming a $\mathbb{Z}_2$-orbit is created in $M$.
	These points
	\begin{equation}\label{qq1}
		\mathbf{q}_*=(q_*, q_*, - q_*, -q_*), \qquad \rho \mathbf{q}_*=-\mathbf{q}_*=(-q_*, -q_*, q_*, q_*)
	\end{equation}
	are $\mathbb{Z}_2\times \mathbb{Z}_2$-symmetric
	because the points of $M$ are fixed by the subgroup
	$\mathbb{Z}_2\times \mathbb{Z}_2=\{e, \rho^2, \kappa, \kappa \rho^2\}$
	of $\mathbb{D}_4$, i.e.\
	the pitchfork bifurcation at $\sigma=\sigma_*$
	%breaks the $\mathbb{D}_4$-symmetry of the zero critical point preserving
	preserves the $\mathbb{Z}_2\times \mathbb{Z}_2$-symmetry of critical points.
	Each point of $M$ in the configuration space corresponds to positioning of the particles in the corners
	of a rectangle. The two rectangles corresponding to critical points $\pm \mathbf{q}_*$ of the potential are mapped to each other by
	the rotation by $\pi/2$, see Figure \ref{figure66:a}.
	
\begin{figure}[h]
	\centering
	\includegraphics*[width=1\columnwidth]{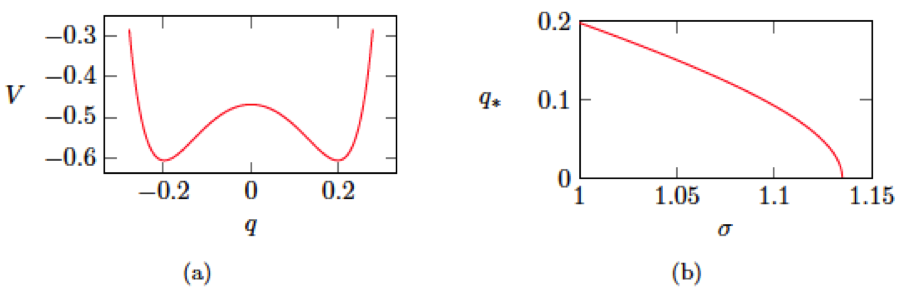}
		\caption{(a) The restriction of potential \eqref{v0} to the subspace $M$ for $\sigma=1$.
		(b) Minimum %\eqref{qq1} corresponding to trapezoid shaped configurations of particles.
		point $q_*$ of the potential shown in panel (a) as a function of $\sigma$.
		\label{figure0':a}\label{figure0':b}\label{figure0'}}
\end{figure}

\begin{figure}[h]
	\centering
\includegraphics*[width=.5\columnwidth]{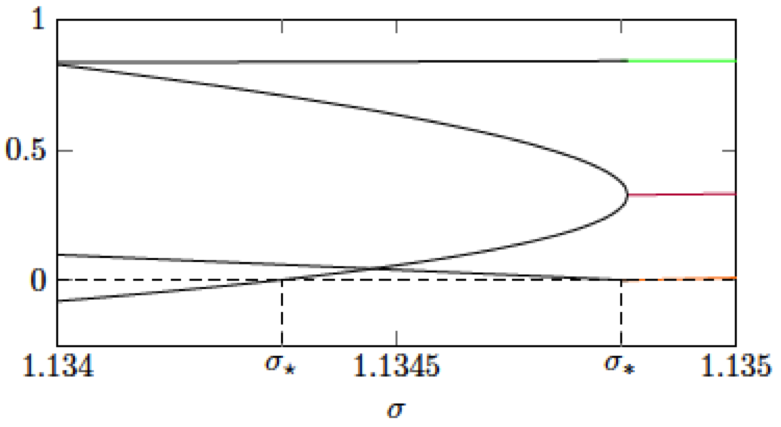}			
\caption{
		Eigenvalues 
		of the Hessian of  potential \eqref{v0} at its critical points \eqref{qq1} as functions of $\sigma<\sigma_*$ (black lines). 
		They merge with the eigenvalues at zero (colored lines)
		at the bifurcation point $\sigma=\sigma_*$. 
		The colors match those in Figure \ref{figure5}.
		\label{figure0}}
\end{figure}

	The components of critical points \eqref{qq1} can be obtained by finding minimum points $\pm q_*$ of the %double-well shaped 
	restriction of $V_0^\sigma$ to $M$:
	\[
	\begin{array}{rcl}
		\hat v_0^\sigma(q)&=&V_0^\sigma(q,q,-q,-q)\ =\ \frac{\sigma^6}{64}\left (-\frac{16}{(q-1)^6} - \frac{16}{(1 + q)^6} - \frac{2}{(1 + q^2)^3}\right.\\
		&+&
		\sigma^6 \left.\left(\frac2{(q-1)^{12}} + \frac2{(1 + q)^{12}} + \frac1{(32 (1 + q^2)^6}\right)\right),
	\end{array}
	\]
	see Figure \ref{figure0':a}. Figure \ref{figure0':b} shows the dependence of $q_*$ on $\sigma$ obtained by numerical minimization (in {\sf Wolfram Mathematica}).
	The eigenvectors of the Hessian on the subspace $M$ are the same as on $L$ and are given by \eqref{eigenvectors}.
	The corresponding eigenvalues of the Hessian on $M$ can be obtained explicitly, they are given by rational expressions in $q$ and $\sigma$. Figure %\ref{figure_trapezium_eig}} 
\ref{figure0}
presents the eigenvalues of the Hessian evaluated at the critical points $\pm \mathbf{q}_*(\sigma)$. %for $\sigma<\sigma_*$.
At the bifurcation value of the parameter, $\sigma=\sigma_*$, they merge with eigenvalues \eqref{lalala} evaluated at the critical point $\mathbf{q}=0$.

One can see that on the interval  \((\sigma_\star,\sigma_*)=\)\((1.13431,1.13483)\), all the eigenvalues at the critical points
$\pm \mathbf{q}_*(\sigma)$ are positive, hence $\pm \mathbf{q}_*(\sigma)$ are minima of the potential $V_0^\sigma$.
However, at the point $\sigma_\star$, one eigenvalue crosses zero, see Figure \ref{figure0}, and these minima destabilize, one in the direction
of the eigenvector $(1,-1,1,-1)$, the other in the direction of the eigenvector $(1,1,1,1)$ (both directions perpendicular to $M$),
hence the critical points $\pm \mathbf{q}_*(\sigma)$ become saddles for $\sigma<\sigma_\star$.
The corresponding two simultaneous pitchfork bifurcations at the points $\pm \mathbf{q}_*(\sigma_\star)$
give rise to a $\mathbb{Z}_4$-orbit of critical points of the potential,
\begin{equation}\label{z4}
	\mathbf{q}_\star, \qquad  \mathbf{q}^\star=\rho \mathbf{q}_\star, \qquad-\mathbf{q}_\star= \rho^2\mathbf{q}_\star, \qquad - \mathbf{q}^\star=\rho^3\mathbf{q}_\star,
\end{equation}
of which $\pm \mathbf{q}_\star$ belong to the two-dimensional subspace of points fixed by the group $\mathbb{Z}_2=\{e, \kappa\}$,
\begin{equation}\label{N0}
	N_0=\{ \mathbf{q}=(q,q,p,p), \ (q,p)\in \mathbb{R}^2\},
\end{equation}
and $\pm \mathbf{q}^\star$ belong to the two-dimensional subspace of points fixed by the group $\mathbb{Z}_2=\{e, \rho^2 \kappa\}$,
\begin{equation}\label{N1}
	N_1=\{ \mathbf{q}=(-q,p,-p,q), \ (q,p)\in \mathbb{R}^2\}.
\end{equation}
A point of $N_0$ corresponds to a configuration of particles forming an isosceles trapezoid, which is symmetric with respect to the line $x=y$;
a point of $N_1$ corresponds to the particles forming a trapezoid, which is symmetric with respect to the line $x=-y$;
and, $\mathbb{Z}_4$-orbit \eqref{z4} corresponds to rotations of an isosceles trapezoid by multiples of $\pi/2$, see Figure \ref{figure66:b}.
Both $N_0$ and $N_1$ are anti-gradient flow invariant.

\begin{figure}[h]
	\centering
\includegraphics*[width=1\columnwidth]{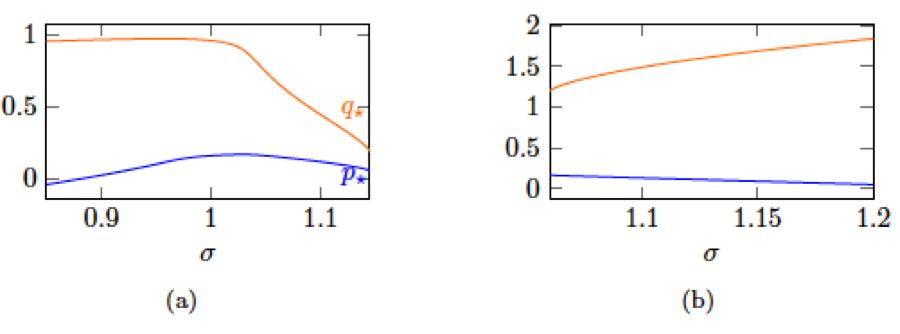}	
	\caption{
		(a) Components
		$q_\star=q_\star(\sigma)$ {(orange)} , $p_\star=p_\star(\sigma)$ {(blue)}
		of a minimum point
		$\mathbf{q}_\star=\mathbf{q}_\star(\sigma)=(q_\star,q_\star,p_\star,p_\star)$ of $V_0^\sigma$
		corresponding to an isosceles trapezoidal configuration of particles {(red nodes on Figure \ref{figure66:b})}. The parameter $\sigma$ ranges over the interval $(.85, 1.145)$ containing the bifurcation point $\sigma_\star$.
		{ (b) Components
			$q_\star(\sigma), p_\star(\sigma)$ of a critical point
			%$\mathbf{q}_\star=\mathbf{q}_\star(\sigma)=(q_\star,q_\star,p_\star,p_\star)$ of $V_0^\sigma$
			corresponding to an isosceles trapezoidal configuration of particles
			with two particles lying outside the square $-1\le x,y\le 1$.
			The range of the parameter $\sigma$ overlaps with that of the branch shown on panel (a).}
	}\label{figure_trapezium:a}
\label{figure_trapezium:b}\label{figure_trapezium}
\end{figure}

\begin{figure}[h]
	\centering
\includegraphics*[width=1\columnwidth]{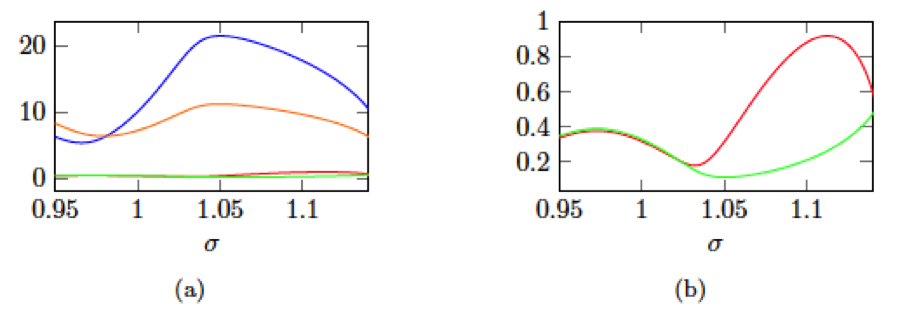}
	\caption{
		(a) Positive eigenvalues of the Hessian for the branch of minimum points shown in Figure \hyperref[figure_trapezium:a]{13a}.
		(b) %shows the 	
		Zoom into the two smaller eigenvalues from panel (a).
	}\label{figure_trapezium_eig}
\end{figure}

\begin{figure}[h]
	\centering
\includegraphics*[width=1\columnwidth]{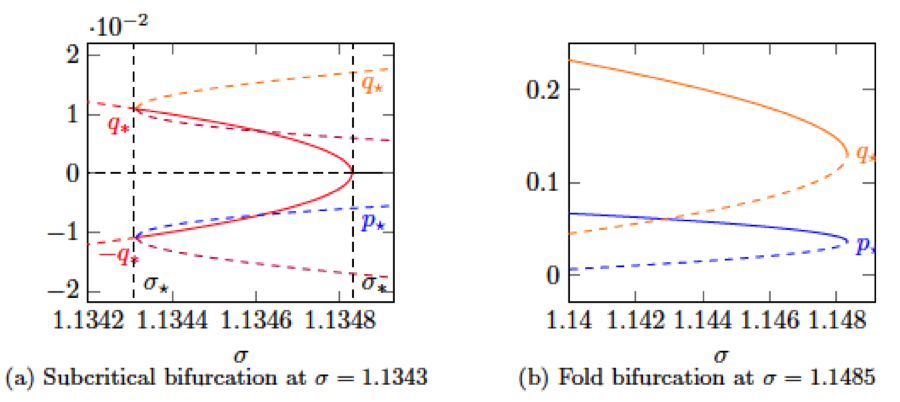}
		\label{figure_s_fold}
	\caption{The branch of trapezoidal configurations shown in Figure \hyperref[figure_trapezium:a]{13a}
		folds at $\sigma=1.485$ (panel (b)) and connects to the branch of rectangular configurations shown in Figure  \ref{figure0':b} at $\sigma=\sigma_\star=1.1343$
		(panel (a)). Solid and dashed segments of the branches correspond to a minimum and a saddle of the potential, respectively.}
\end{figure}

Minimizing $V_0^\sigma$ on $N_0$ provides the branch
of critical points $\mathbf q_\star(\sigma)$. We restrict our attention to the segment of this branch shown in Figure \hyperref[figure_trapezium:a]{13a}
with $\sigma$ ranging over the interval $(.95, 1.145)$, which contains the bifurcation point $\sigma_\star$.
The eigenvalues of the Hessian are positive on this segment (see Figure \ref{figure_trapezium_eig}), hence $\mathbf q_\star(\sigma)$ is a minimum point of the potential,
and so are all four points of $\mathbb{Z}_4$-orbit \eqref{z4}.
The branch containing this segment connects to the the branch of rectangular configurations \eqref{qq1} via the fold bifurcation
at $\sigma=1.148$, see
Figure \ref{figure_s_fold} and the subcritical bifurcation at the point $\sigma_\star$, see
Figure \ref{figure_s_subcritical}.

It is worth noting that on the parameter interval {$1.1384<\sigma < 1.145$} the minimum at zero co-exists with four minimum points \eqref{z4} creating muli-stability for $h=0$.

\begin{figure}[h]
	\centering
	\includegraphics*[width=1\columnwidth]{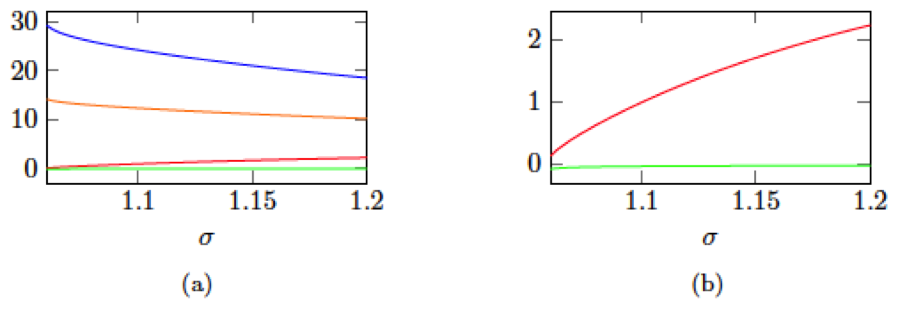}
	\caption{
		(a) Eigenvalues of the Hessian for the branch of critical points shown in Figure \ref{figure_trapezium:b}.
		Panel (b) zooms into the two smaller eigenvalues from panel (a). One eigenvalue is negative, i.e.\ the critical points are saddles.
	}\label{figure_bigbranch_eig}
\end{figure}

Branches of minima \eqref{z4} shown in Figure \hyperref[figure_trapezium:a]{13a} correspond to isosceles trapezoidal configurations of particles located within the square $-1\le x,y \le 1$ (equivalently, $-1\le q, p\le 1$).
In addition, the potential $V_0^\sigma$ has a $\mathbb{Z}_4$-orbit of critical points which also belong to the subspaces $N_0$, $N_1$
but correspond to isosceles trapezoidal configurations with two particles located outside the square $-1\le q,p \le 1$, see Figure \ref{figure_trapezium:b}.
These are saddle points with one unstable direction which is perpendicular to the subspace $N_0$  (resp., $N_1$) where the critical point is located, see Figure \ref{figure_bigbranch_eig}. There is an interval of the parameter $\sigma$ within which these critical points co-exist with the minimum points
shown in Figure \hyperref[figure_trapezium:a]{13a}.
We notice that the restriction of the potential to the subspace $N_0$,
\[
\begin{array}{rcl}
	V_0^\sigma(q,q,p,p)&=&\frac{\sigma^6}{64} \left(-\frac{8}{(q-1)^6} - \frac{128}{(4 + (q - p)^2)^3} -
	\frac{128}{((1 + q)^2 + (p-1)^2)^3} - \frac{8}{(1 + p)^6}
	\right.\\
	&+ &\left.
	\sigma^6 \left (\frac{1}{(q-1)^{12}} + \frac{128}{(4 + (q - p)^2)^6}
	+ \frac{128}{((1 + q)^2 + (p-1)^2)^6} + \frac{1}{(1 + p)^{12}}\right)
	\right),
\end{array}
\]
has a singularity on the lines $q=1$, $p=-1$ and at the point $(q,p)=(-1,1)$.
The restriction of this potential to the subspace $N_1$ has singuarities at the same locations.

In Section \ref{expansion}, we will consider the system under expansion (potential \eqref{exp}) for $\sigma<\sigma_{*}$ and show hysteresis between isosceles trapezoidal configurations of particles as the external forcing parameter $h$ is varied.

\subsection{System under rotational forcing}\label{rotation}

Let us consider potential \eqref{rot} with rotational external forcing for $\sigma^*<\sigma <\sigma^{**}$ (cf.\ \eqref{addd}). As shown in the previous subsection, when the forcing is zero ($h=0$), the potential has two minimum points \eqref{qq} corresponding to square-shaped configurations of particles shown in Figure \ref{figure65}.

Potential \eqref{rot} is invariant with respect to the action of the group $\mathbb{Z}_4$ but not invariant with respect to the action
of the group $\mathbb{Z}_2$. However, we observe that
\[
V_h^\sigma(\mathbf{q})= V_{-h}^\sigma(\kappa\mathbf{q}).
\]
In particular, if $\mathbf{q}$ is a local minimum point of the potential for some $h$, then $\kappa \mathbf{q}$ is a local minimum point for the value $-h$ of the forcing parameter.

\begin{figure}[h]
	\centering
\includegraphics*[width=.9\columnwidth]{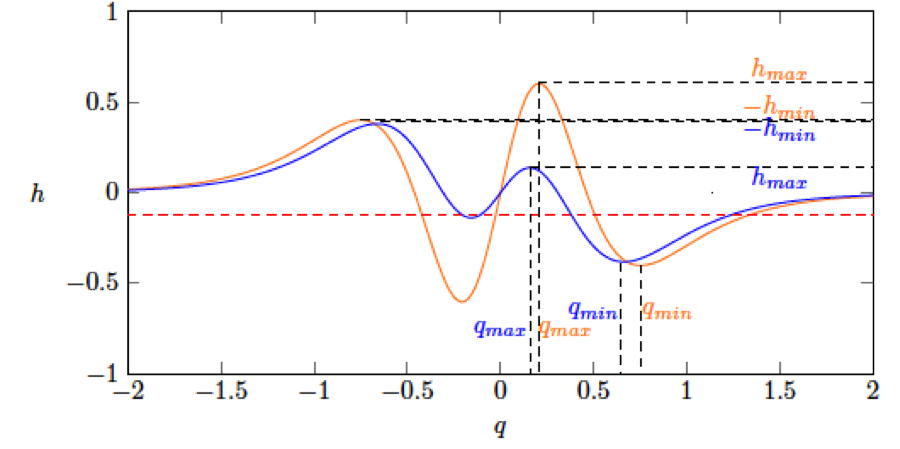}
	\caption{Plot of the function $-(v_0^\sigma )'(q)/4$ for $\sigma=1.33$ (blue) and $\sigma=1.4$ (orange).
		Intersections of the graph with a horizontal line $y=h$ define critical points $q^*(h)$ of the function $v_h^\sigma(q)$.
	}\label{figure7}
\end{figure}

\begin{figure}[ht!]
\includegraphics*[width=.9\columnwidth]{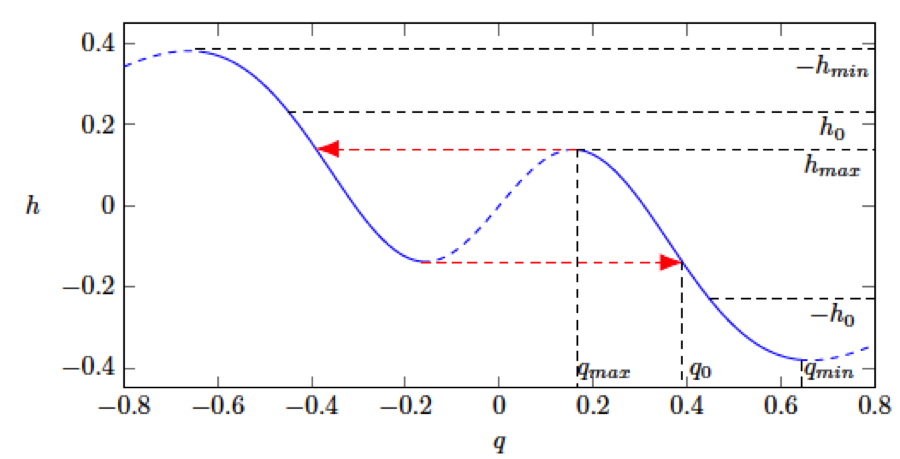}
	\caption{
		Hysteresis loop for potential \eqref{rot} with external forcing $h$. The system moves along the straight line $L$ of $\mathbb{Z}_4$-symmetric states, hence the position in the configuration space is described by one scalar parameter $q$.
		The {blue} %dashed 
		curve is the graph of the function $-(v_0^\sigma )'(q)/4$, see the blue curve in Figure \ref{figure7}. Solid parts of the curve correspond to minimum points of the potential.
		As $h$ increases from the minimal value $-h_0$, the point $(q^*(h),h)$ follows the solid part %blue segment 
		on the right branch of the %dashed red 
		curve %from right to left the rightmost lowest point of the blue segment 
		moving upwards left in the direction of the local maximum of the %red 
		curve. In the configuration space, the system sits at the (moving) local minimum point $\mathbf{q}^*(h)=(q^*(h), -q^*(h), -q^*(h), q^*(h))$ of the potential. Once the point $(q^*(h),h)$ reaches the maximum point of the %red 
		curve at $h=h_{max}$, it transits horizontally along the dashed 
		%blue 
		arrow to the left branch of the %red 
		curve, which corresponds to the local minimum point $-\mathbf{q}^*(-h)=\kappa \mathbf{q}^*(-h)$ of the potential. In the configuration space,  this event corresponds to the local minimum $\mathbf{q}^*(h)$ disappearing in the saddle-node bifurcation at $h=h_{max}$, and the system transitioning to the remaining minimum point $-\mathbf{q}^*(-h)$ along the line $L$. Now, as $h$ increases further, the point $(q^*(h),h)$ follows the %blue 
		solid segment on the left branch of the %dashed red 
		curve until it reaches the %leftmost 
		highest point %of this segment 
		at $h=h_0$. Similarly, as $h$ decreases from the maximum value $h_0$, the point $(q^*(h),h)$ follows the left branch of the 
		%red 
		curve downwards right, transits along the horizontal dashed %blue 
		%segment 
		arrow to the right branch of the %red 
		curve at the point $h=-h_{max}$, and continues along the right branch until it reaches the rightmost lowest 
		%end of the blue solid segment 
		point at $h=-h_0$.
		In the configuration space, the system sits in the local minimum point $-\mathbf{q}^*(-h)$ until this minimum disappears in the saddle-node bifurcation at $h=-h_{max}$, at which point the system transitions to the local minimum $\mathbf{q}^*(h)$ along the line $L$, and then remains at $\mathbf{q}^*(h)$ until $h$ reaches the value $-h_0$. 
	}\label{figure8}
\end{figure}

As in the case without forcing, subspace \eqref{L} of $\mathbb{Z}_4$-symmetric points is invariant for the anti-gradient flow of potential \eqref{rot}.
Restricting the potential to $L$, we obtain the scalar function
\[
v^\sigma_h(q)= 4h q + v_0^\sigma (q)=4h q+\frac{ 129 \sigma^{12} - 1088 \sigma^6(1 + q^2)^3}{2048 (1 + q^2)^6},
\]
whose critical points $q^*$ define $\mathbb{Z}_4$-symmetric critical points $\mathbf{q}^*=(q^*,-q^*,-q^*,q^*)$ of the potential $V_h^\sigma$.
Hence, we consider zeros of the derivative
\[
-\frac{(v^\sigma_h)'(q)}4=-\frac{(v_0^\sigma )'(q)}4-h=\frac{3 s^6 q (129 s^6 - 544 (1 + q^2)^3)}{2048 (1 + q^2)^7}-h.
\]
The graph of the odd function $-(v_0^\sigma )'/4$ has four extremum points for $\sigma>\sigma^*$, see Figure \ref{figure7}.
In particular, on the positive semi-axis the local maximum and minimum points satisfy
\begin{equation}
	0< q_{max} < q_{min}, \quad h_{max} := -\frac{(v_0^\sigma )'(q_{max})}4 >0 > -\frac{(v_0^\sigma )'(q_{min})}4=:h_{min}.
\end{equation}
Therefore,
hysteresis occurs if the following conditions are satisfied:
\begin{itemize}
	
	\item The local extremum values of the function $-(v_0^\sigma)'/4$ satisfy $h_{max}<-h_{min}$, see the blue plot in Figure \ref{figure7} (the orange plot violates this condition).
	From this condition, it follows that there is a unique point $q_0$ satisfying
	\begin{equation}\label{q0}
		q_{max}<q_0<q_{min}, \qquad %such that
		-\frac{(v_0^\sigma)'(q_0)}4 =
		-\frac{(v_0^\sigma )'(-q_{max})}4 =
		-h_{max},
	\end{equation}
	see Figure \ref{figure8}.
	
	\item  Assuming that the external forcing parameter $h$ oscillates between $-h_0$ and $h_0$, the amplitude $h_0$ satisfies $h_{max} < h_0 < - h_{min}$.
	
	\item The segment of the straight line \eqref{L} between the points $\pm (q_0,-q_0,-q_0,q_0)$ is transversally stable. In other words, the eigenvalues
	\begin{equation}\label{l1l2}
		\lambda_1=\lambda_2=
		\frac{9 \sigma^6 \bigl(\sigma^6 - 2 (1 + q^2)^3\bigr)}{8 (1 +
			q^2)^7},
	\end{equation}
	\begin{equation}\label{l3}
		\lambda_3=-\frac{
			3 \sigma^6 \bigl(-96 (1 + q^2)^3 (-37 + 3 q^2) + \sigma^6 (-1663 + 115 q^2)\bigr)
		}
		{2048 (1 + q^2)^8}
	\end{equation}
	of the Hessian, which correspond to the eigenvectors $(0, 1, 0, 1),$ $(1, 0, 1, 0),$ $(-1, -1, 1, 1)$ orthogonal to $L$ (see \eqref{eigenvectors}),
	are positive on the segment $-q_0\le q\le q_0$.
	%\textcolor{red}{
		We note that the eigenvalue
		\[
		\lambda_4=\frac{3 \sigma^6 \bigl(-544 (1 + q^2)^3 (-1 + 7 q^2) +
			129 \sigma^6 (-1 + 13 q^2)\bigr)}{2048 (1 + q^2)^8}=\frac{(v_F^\sigma)''(q)}4
		\]
		corresponding to the eigenvector $(1,-1,-1,1)$ in the direction of $L$ is negative on the interval $(-q_{max},q_{max})$
		and positive on each of the intervals $(-q_{min},-q_{max})$ and $(q_{max},q_{min})$ which include the points $-q_0$ and $q_0$, respectively.
		%}
\end{itemize}

Under these conditions, the system with potential \eqref{rot} exhibits hysteresis as shown in Figure \ref{figure8}.

\begin{figure}[h]
	\centering
	\includegraphics*[width=1\columnwidth]{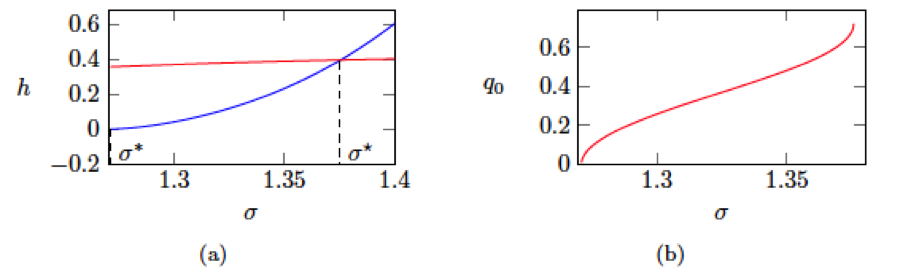}
		\label{figure9}
		\label{figure10}
	\caption{
		(a) Dependence of the the extremum values $h_{max}$ (blue) and $-h_{min}$ (red) of the function $-(v_0^\sigma )'/4$ on $\sigma$.
		The value of $\sigma$ at the intersection point is $\sigma^\star=1.375$. %1.37556778
		(b) Dependence of $q_0$ on $\sigma$ on the interval $(\sigma^*,\sigma^\star)$.
	}
\end{figure}

The first of the above three conditions is satisfied for the values of $\sigma$ from the interval $(\sigma^*,\sigma^\star)=(1.27107,1.375)$, %1.3755
see Figure \ref{figure9} which shows the dependence of $h_{max}$ and $-h_{min}$ on $\sigma$. The second condition is satisfied for every pair $(\sigma,h_0)$ in the region bounded above by the graph of $-h_{min}(\sigma)$ (red line) and below by the graph of $h_{max}(\sigma)$ (blue line) on the same figure.

\begin{figure}[ht!]
	\centering
\includegraphics*[width=.8\columnwidth]{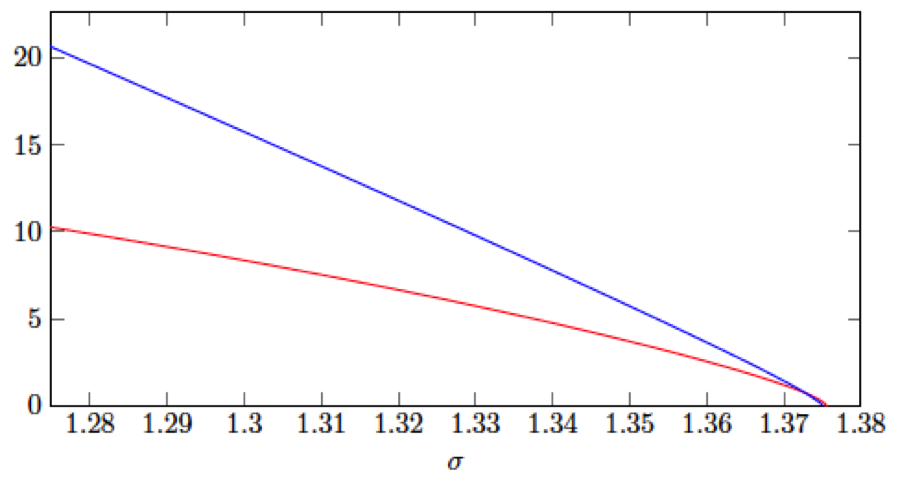}
	\caption{
		Transversal eigenvalues $\lambda_1=\lambda_2$ (red) and $\lambda_3$ (blue) evaluated at the point $q_0(\sigma)$ as functions of $\sigma$ on the interval $(\sigma^*,\sigma^\star)$.
	}\label{figure11}
\end{figure}

The third condition involves the interval $(-q_0,q_0)$ where $q_0$ is defined non-locally by equation \eqref{q0} (see Figure \ref{figure8}).
Figure \ref{figure10} shows the dependence of $q_0$ on $\sigma$ on the interval of interest, $(\sigma^*,\sigma^\star)$.
As confirmed by Figure \ref{figure11}, the
eigenvalues $\lambda_1=\lambda_2$, $\lambda_3$ (see \eqref{l1l2}, \eqref{l3}) evaluated at $q=q_0(\sigma)$ are positive for the values of $\sigma$ from this interval. %$\sigma^* <\sigma < \sigma^\star$, see Figure \ref{figure11}.
These eigenvalues are even functions of $q$, the eigenvalues $\lambda_1=\lambda_2$ decrease with $q$ for $q\ge0$, and the eigenvalue $\lambda_3$ also decreases with $q$
in the domain of interest, i.e.\ in
\begin{equation}\label{domain}
	\bigl\{ (\sigma,q): \sigma^*< \sigma < \sigma^\star, \ 0 \le q \le q_0(\sigma) \bigr\}.
\end{equation}
Hence, Figure \ref{figure11} %(a,b) 
ensures that all the transversal eigenvalues are positive on the segment $-q_0(\sigma)\le q\le q_0(\sigma)$ for each $\sigma$ from the interval
$(\sigma^*,\sigma^\star)$, i.e.\ the third condition is also satisfied on this interval.

Hence, we conclude that the system with potential \eqref{rot} exhibits hysteresis if the parameter $\sigma$ of the potential satisfies $\sigma^*< \sigma < \sigma^\star$. It is the same type of hysteresis associated with bi-stability as shown in Figure \ref{fb2}.

Clearly, the symmetric range of $h$ can be replaced by any asymmetric range $h_1 \le h\le h_2$ provided that $ h_{max} < -h_1, h_2< -h_{min}$.

\subsection{System under expansion}\label{expansion}

In this section, we consider potential \eqref{exp} for the fixed $\sigma=1.12<\sigma_*$ and vary the force parameter $h$. This potential is invariant with respect to the action of the subgroup $\mathbb{Z}_2\times \mathbb{Z}_2=\{e, \rho^2, \kappa, \kappa \rho^2\}$
of $\mathbb{D}_4$
and satisfies
\[
\hat V_h^\sigma(\mathbf{q})= \hat V_{-h}^\sigma(\rho\mathbf{q}).
\]
Therefore, the planes $N_0, N_1$ defined by \eqref{N0}, \eqref{N1} (which correspond to
isosceles trapezoidal formations of particles, see Figure \ref{figure66:b})
are invariant for the gradient field in the configuration space.

\begin{figure}[ht!]
	\centering
\includegraphics*[width=.8\columnwidth]{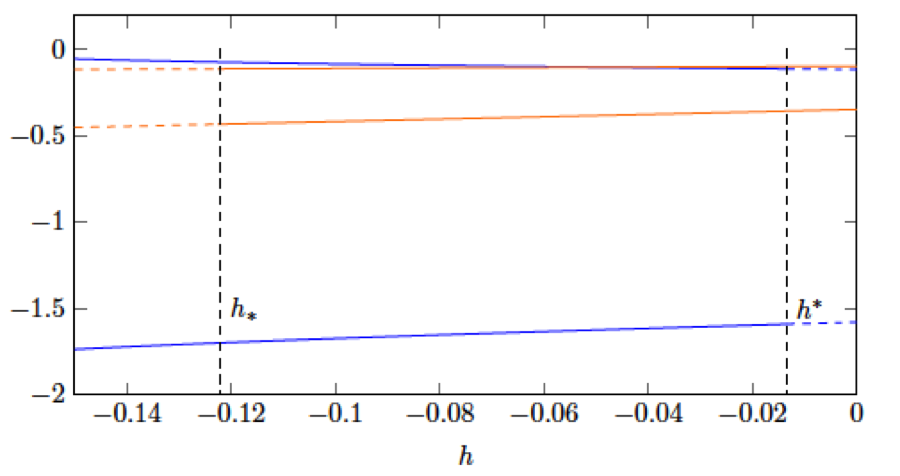}
	\caption{
		Components $q,p$ for two branches of isosceles trapezoidal critical points
		$(q,q,p,p)\in N_0$. For the {yellow} branch, the formation of particles belongs to the square $-1\le x,y\le 1$; for the {blue} branch, two particles are located outside this square; $-0.15<h<0$; $\sigma=1.12$. 
	}\label{figure_SF}
\end{figure}

\begin{figure}[h]
	\centering
\includegraphics*[width=1\columnwidth]{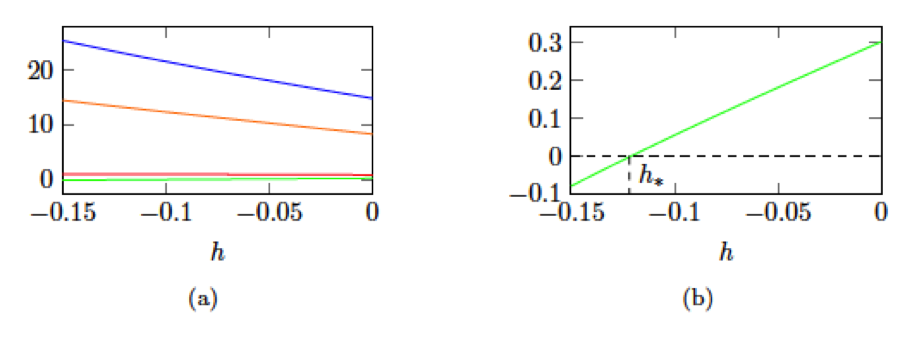}
	\caption{
		(a) Eigenvalues for the {yellow} branch of critical points shown in Figure \ref{figure_SF}. (b) The smallest eigenvalue corresponding to a direction perpendicular to $N_0$. 
	}\label{figure_F_small_eig}
\end{figure}

\begin{figure}[h]
	\centering
	\includegraphics*[width=1\columnwidth]{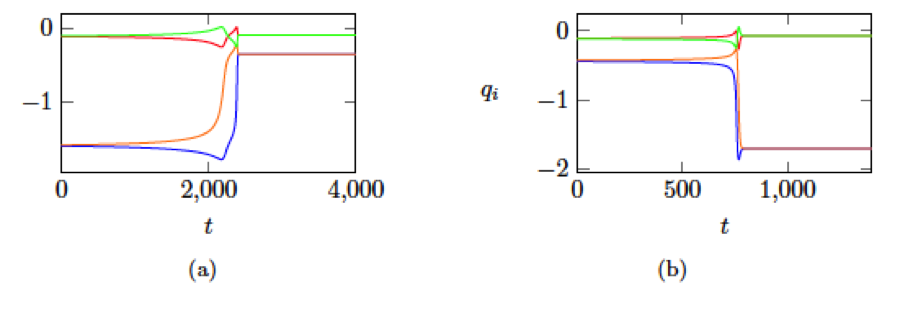}	
	\label{figure_transition:b}
		\label{figure_transition:a}
	\caption{
		(a) Transition from the {yellow} branch to the {blue} branch at the bifurcation point $h=-0.122$. (b) The backward transition at the bifurcation point $h=-0.0135
		$. Each panel shows the time plots of the coordinates $q_i$ of $\mathbf q$ during the corresponding transition, which follows the anti-gradient dynamics $\dot {\mathbf q}=-\nabla \hat V_h^\sigma({\mathbf q})$.  }
\label{figure_transition}
\end{figure}

Figure \ref{figure_SF} presents two branches of critical points located in the plane $N_0$.
Eigenvalues along the yellow branch are shown in Figure \ref{figure_F_small_eig}. This critical point is a minimum for $h=0$.
As $h$ decreases, the smallest eigenvalue becomes negative at $h=h_*=-0.122$.
The corresponding saddle-node critical point is
\[
(q_1,q_2,q_3,q_4)=(-0.430867, -0.430867, -0.110452, -0.110452).
\]
Figure \ref{figure_transition:a} shows the transition from the above critical point to the minimum point
\[
(q_1,q_2,q_3,q_4)=(-1.69683, -1.69683, -0.0720405, -0.0720405)
\]
on the {blue} branch resulting from a small perturbation in a direction perpendicular to the subspace $N_0$ of isosceles trapezoidal configurations.

\begin{figure}[ht!]
	\centering
	\includegraphics*[width=1\columnwidth]{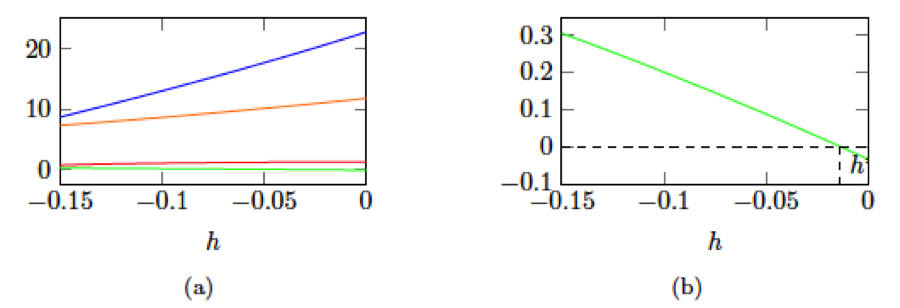}
	\caption{
		(a) Eigenvalues for the {blue} branch of critical points shown in Figure \ref{figure_SF}. (b) The smallest eigenvalue corresponding to a direction perpendicular to $N_0$.
	}\label{figure_F_big_eig}
\end{figure}

\begin{figure}[h]
	\centering
\includegraphics*[width=1\columnwidth]{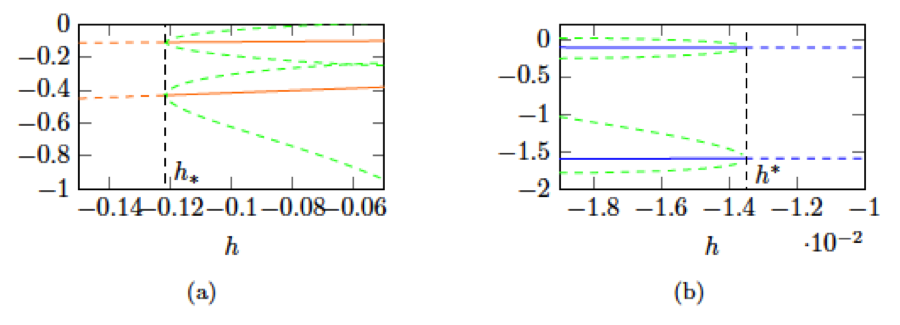}
	\caption{Subcritical $\mathbb{Z}_2$-symmetry breaking pitchfork bifurcations at the points (a) $h_*=-0.122$ and (b) $h^*=-0.0135$, which limit the bi-stability range
		of the $\mathbb{Z}_2$-symmetric trapezoidal solutions.
	}\label{figure_subcritical}
\end{figure}

Eigenvalues along the {blue} branch are shown in Figure \ref{figure_F_big_eig}. As $h$ increases from the value $h_*$, the smallest eigenvalue becomes negative at $h=h^*=-0.0135$.
The corresponding critical point is
\[
(q_1,q_2,q_3,q_4)=(-1.58907, -1.58907, -0.110771, -0.110771).
\]
Figure \ref{figure_transition:b} shows the backward transition from this saddle-node critical point to the minimum point
\[
(q_1,q_2,q_3,q_4)=(-0.355008, -0.355008, -0.0969899, -0.0969899)
\]
on the {yellow} branch resulting from a small perturbation in a direction perpendicular to the subspace of isosceles trapezoidal configurations $N_0$.
Thus, varying $h$ over an interval $[h_0,h^0]$ which satisfies
$%[-0.122 , -0.135]
[h_*,h^*]\subset [h_0,h^0] \subset [-0.15,0]$ results in a hysteresis loop.
Bifurcations at the points $h=h_*,h^*$ are subcritical pitchfork bifurcations associated with $\mathbb{Z}_2$-symmetry breaking of the isosceles trapezoidal solutions, see Figure \ref{figure_subcritical}. %Figures \ref{figure_subcriticaleig}, \ref{figure_subcriticaleig_big} show the eigenvalues of the Hessian along the branches of non-symmetric saddle points represented by the parabolas in Figure \ref{figure_subcritical}.

\section{Conclusions}
A particle in a quasistatically varied double-well potential is a canonical example of hysteresis associated with bi-stability and elimination of a minimum of the potential energy via a saddle-node bifurcation. We explored similar scenarios in systems of particles assuming a natural single-well (Lennard-Jones) interaction potential for each pair of particles. In this setting, if $N$ identical particles are constrained to a straight line, each particle interacts with its nearest neighbors, and a quasistatically varied external forcing is applied at the ends of the chain, then the potential energy has at most one minimum, hence the system doesn't exhibit hysteresis. Therefore, we considered particles on the plane.
Two hysteresis scenarios were shown in a simple (constrained) four-particle system with four degrees of freedom. The first scenario is equivalent to a one-degree-of-freedom particle in a double-well potential because the %anti-gradient
evolution in the configuration space is restricted to a one-dimensional invariant attracting manifold (straight line) of symmetric square-shaped configurations.
In the second scenario, critical points of the potential which are restricted to an invariant plane of isosceles trapezoidal configurations are destabilized by a symmetry breaking bifurcation in a transversal direction, hence the ensuing transient dynamics towards a minimum occurs outside the plane where the minima are located.
% through four dimensions.

Important phenomenological models of hysteresis (such as models of constitutive relations of materials and media) combine, or admit decomposition into, many bi-stable elements. As such, they exhibit specific types of hysteresis, which can be identified by properties of hysteresis loops. For example, hysteresis loops of the Ising, Preisach and Prandtl-Ishlinskii models close after one period (the so-called return point memory property); additionally, all hysteresis loops of the Preisach model corresponding to the same periodic input are congruent to each other; all the loops of the Prandtl-Ishlinskii model are centrally symmetric. It would be interesting to characterize hysteresis of multi-particle systems, in which particles interact via the Lennard-Jones potential (as in \eqref{inter}--\eqref{vdw}), and compare it to the types of hysteresis exhibited by standard phenomenological models.
One particular example of such multi-particle systems are amorphous media, %especially 
specifically low-molecular and polymer glasses. Plastic phenomena in these systems are closely related to the succession of bifurcations of their complicated multi-dimensional potential landscape \cite{ML,GP}. Numeric simulations, both in athermal quasistatic regime and with molecular dynamics, demonstrate clear hysteretic behavior, in complete agreement with physical intuitive apprehension of plasticity. Still, a direct relationship between this hysteresis and the particularities of interatomic interactions remains mysterious.
However, these %analysis
questions
are beyond the scope of this work. It would be also interesting to replace transitions along the anti-gradient field with inertial transition dynamics $m \ddot {\bf q} + \gamma \dot {\mathbf q} + \nabla V({\mathbf q};h) = 0$. The anti-gradient transitions correspond to the limit of large friction forces. In the opposite frictionless limit (i.e., $m \ddot {\bf q} + \nabla V({\mathbf q};h) = 0$), transitions are initiated by saddle-center bifurcations and end at oscillating regimes.

It is worth noting that any type of hysteresis is possible in a two-degrees-of-freedom system if the class of potentials is not restricted.
% (e.g., to Lennard-Jones interactions as in this work, or non-interacting double-well potentials as in the case of the Preisach model).
To make this statement precise,
%hysteresis was described in \cite{} as a directed graph, where
an edge-labeled directed graph $\Gamma$ was associated in \cite{rachinskii} with any $N$-degree-of-freedom potential energy $V_h({\mathbf q})$ as follows.
With each energy minimum (state)
%${\mathbf q}_j={\mathbf q}_j(h)$
that exists on an input interval
\begin{equation}\label{l}
	h^-_j < h < h^+_j
\end{equation}
(where $h^\pm_j$ are saddle-node bifurcation points), one associates a graph vertex $v_j$. Every vertex has two outgoing directed edges.
One edge, labeled $h_j^-$, corresponds to the transition from the state labeled $v_j$ to another state as a decreasing input $h$ reaches the bifurcation value $h_j^-$; the other edge, labeled $h_j^+$, corresponds to the transition, which occurs when an increasing input reaches the bifurcation value $h_j^+$.
Since the graph $\Gamma$ %creates a `hysteresis map' of all transitions between states in response to quasistatic variations of the onput.
encodes all the transitions between states in response to quasistatic variations of the input,
it is called a hysteresis map for $V_h$.
By design, for any vertex $v_j$, the labels $h$ of all the incoming edges satisfy \eqref{l}.
As shown in \cite{rachinskii}, {\em any} edge-labeled directed graph $\Gamma$ which, at each vertex, has exactly two outgoing edges, with the incoming edge labels $h$ and outgoing edge labels $h_j^\pm$ satisfying \eqref{l}, is a hysteresis map for some two-degrees-of-freedom potential $V_h(q_1,q_2)$.
It would be interesting to determine what hysteresis maps correspond to multi-particle potentials \eqref{inter} with Lennard-Jones interactions.

\section*{Acknowledgments}
This work was supported by Lady Davis Visiting Professorship at Technion---Israel Institute of Technology.

\end{document}